\documentclass[11pt]{article}

\usepackage{amsmath,amssymb}
\usepackage{slashed}
\usepackage{xcolor} 
\usepackage{graphicx}
\usepackage{url}

\newcommand{\be}{\begin{equation}}
\newcommand{\ee}{\end{equation}}
\newcommand{\bea}{\begin{eqnarray}}
\newcommand{\eea}{\end{eqnarray}}
\def\Re{{\cal R \mskip-4mu \lower.1ex \hbox{\it e}\,}}
\def\Im{{\cal I \mskip-5mu \lower.1ex \hbox{\it m}\,}}

\def\tev{\,{\ifmmode\mathrm {TeV}\else TeV\fi}}
\def\gev{\,{\ifmmode\mathrm {GeV}\else GeV\fi}}
\def\mev{\,{\ifmmode\mathrm {MeV}\else MeV\fi}}
\def\to{\rightarrow}

\begin{document}

\vspace{4cm}
\begin{center}
{\Large\bf{Direct photon production as a probe of quarks chromoelectric and chromomagnetic dipole moments at the LHC
}}\\

\vspace{1cm}
 {\bf Hoda Hesari and Mojtaba Mohammadi Najafabadi }  \\
\vspace{0.5cm}
{\sl  School of Particles and Accelerators, \\
Institute for Research in Fundamental Sciences (IPM) \\
P.O. Box 19395-5531, Tehran, Iran}\\

\end{center}
\vspace{2cm}
\begin{abstract}
In this paper we show that the $\gamma+$jet invariant mass distribution
in proton-proton collisions at the LHC is significantly sensitive to the quarks chromoelectric ($\kappa$) and chromomagnetic 
($\tilde{\kappa}$) dipole moments. It is shown that the presence of $\kappa$ or $\tilde{\kappa}$
leads to an increment of the cross section of $\gamma+$jet process in particular in the tail of 
$\gamma+$jet invariant mass distribution.
Using the measured $\gamma+$jet invariant mass distribution by the CMS experiment at the center-of-mass energy
of 8 TeV, we derive bounds on the quarks chromoelectric and chromomagnetic dipole moments.
In extraction of the limits, we consider both theoretical and systematic uncertainties. The uncertainties originating from 
variation of the renomalization/factorization scales and the choice of proton parton distribution
functions are taken into account as a function of the $\gamma+$jet invariant mass. We exclude $\kappa$ or $\tilde{\kappa}$ above
$10^{-5}$ at 95$\%$ confidence level. This is the most stringent direct upper limit on $\kappa$ or $\tilde{\kappa}$.
\end{abstract}
\vspace{1cm}
PACS number(s):13.40.Em, 13.85.Qk

\newpage

\section{Introduction}

Standard Model (SM) is a framework that describes our present understanding of
fundamental constituents of matter and their interactions. 
The SM predictions are found to be well compatible with the experimental data up to the 
scale a few TeV. While already the information which have come out of the LHC and Tevatron experiments
show consistency with  the SM expectations with good precisions, the LHC future runs provide the possibility to 
achieve more precise determination of the SM particle properties. 
Although no significant indication for new physics beyond the
SM has been found up to now, more precise measurements of
SM predictions can reveal tracks of the effects of new physics beyond the SM.
A systematic and powerful way to parametrize the new physics effects is to utilize the
effective Lagrangian approach which is a model independent tenchique in the probe of new physics effects.
In this approach, the effective Lagrangian follows the same symmetries as the SM
and is constructed from the existing SM fields. The leading terms
in $\mathcal{L}_{eff}$ are the SM terms and the effects of new physics is parametrized by the coefficients of higher dimension
operators. The effective Lagrangian for exploring the new interactions has been provided in \cite{wb,gimr}.
The lowest order couplings between gluon and a quark are dimension four and five operators with the 
following form:
\begin{eqnarray}
\label{eff}
{\cal L}_{eff} = g_{s}T^{a}\bar q \left[-\gamma^{\mu}G^{a}_{\mu}+\frac{\kappa}{4m_q} 
\sigma^{\mu\nu} G_{\mu\nu}^a - \frac{i \tilde{\kappa}}{4m_q} \sigma^{\mu\nu} \gamma^5 G_{\mu\nu}^a \right]q,
\end{eqnarray}  
where $G_{\mu}^{a}$ denotes the gluon field, $\kappa/2m_{q}$ and $\tilde{\kappa}/2m_{q}$ 
are corresponding to chromomagnetic (CMDM) and chromoelectric (CEDM) dipole moments of a quark.
It should be noted that within the SM these couplings are zero at
tree level and are induced at loop level. 
The chromoelectric and chromomagnetic dipole moments $\kappa$ and $\tilde{\kappa}$
can be  considerable since they appear as dimension five operators and are only suppressed by 
one power of $\Lambda$ (a new physics scale). 
In the denominator the mass factors are taken conventionally to be the quark mass $m_{q}$
to express these terms as quark dipole moments. 

The effective Lagrangian $\mathcal{L}_{eff}$ is valid for all quark flavors within the vast range of quarks masses.
There is much interest to measure CEDM since a nonzero value
of CEDM indicates a new source of CP violation. 
In the SM, the amount of quarks chromoelectric moment (CEDM) are very small. For example,
the CEDM of the heaviest quark i.e. top quark is at the order of $10^{-17}$ $g_{s}.$cm \cite{sm}.
Several extensions of the SM such as Minimal Supersymmetric Standard Model (MSSM), GUT theories, 
Two Higgs Doublet Model, Higgs Triplet Model, Left-Right Symmetric Model and Extra Dimensions
can generate sizeable (chromo) electric  and (chromo) magnetic dipole moments \cite{fDM,mpp,lc,dma,ho,13,14}. 

So far, there are several studies on constrainig the top quark CEDM and CMDM in the litrature using 
different methods. In \cite{tw}, we have used single top in $tW$ channel to probe the dipole moments.
In \cite{tt1,tt2,tt3,tt4,tt5,tt6,tt7,tt8,tt9,tt10}, the total and the differential cross sections of top pair production
at the LHC and Tevatron have been used to constrain the dipole moments.

In hadron colliders, direct photon production ($pp(\bar{p})\rightarrow \gamma+jet$)
provides very useful information to search for new physics beyond the SM as well as
increasing our knowledge of SM. For example, within the SM the total and differential $\gamma+jet$ 
cross sections are used to understand the proton parton distribution
functions (PDFs) and even are used for testing the perturbative QCD \cite{ketall,fh,hs}. 
From the experimental point of view $\gamma+jet$ events are clean and
are well measured with the electromagnetic calorimeters which are used in 
improving photon energy resolution. 
More importantly, $\gamma+jet$ is the final state of many new physics signatures such as
quantum black holes (QBHs) \cite{1,2,3}, excited quarks \cite{4,5,6}, quirks \cite{7,8,9} and Regge excitations of string theory \cite{10,11,12}.
Indeed, this is not the complete list of new physics models that can be explored with photons at hadron colliders 
however this shows the ability of photon final state which covers a broad range of
theoretical models beyond the SM.  The results of the experimental searches based on 
$\gamma+jet$ final state at the LHC and Tevatron can be found in \cite{fabe,kd,cms,atlas1,atlas2,atlas3,atlas4,atlas5}.

In \cite{kd}, the authors have studied the influence of anomalous
CMDM ($\kappa$) and CEDM  $(\tilde{\kappa})$ of light quarks on the direct photon production
at the Tevatron. It has been shown that the $\gamma+jet$ rate is sensitive
to anomalous interactions of quarks to gluons in particular the differential cross section $d\sigma/dp_{T\gamma}$.
The transverse momentum spectrum has been found to be sensitive to CEDM and CMDM. 
Nonzero values of CEDM or CMDM enhances the cross section in the photon high transverse momentum region. 
The upper bound of 0.0027  has been set on $\kappa$ using 0.1 fb$^{-1}$ of CDF and D0 data.

In this paper, we show that the presence of anomalous CEDM or CMDM of quarks 
increases the rate of photon production in the tail
of $\gamma+jet$ invariant mass distribution at the LHC. Then using the recent $\gamma+jet$
spectrum measurement at the center-of-mass energy of 8 TeV with 19.7 fb$^{-1}$ of data, upper 
bounds are set on CEDM and CMDM at $95\%$ confidence level.
In limit setting process, a special attention is paid to the uncertainties.
We calculate the uncertainty originating from variation of renormalization/factorization scales
and the uncertainty coming from the limited knowledge of proton parton density functions (PDFs)
as functions of the  $\gamma+jet$ invariant mass.

This paper is organized as follows. Section 2 is dedicated to theoretical calculations of 
the cross section of $\gamma+jet$.  Section 3 describes the sensitivity estimate using the
measured mass spectrum of photon-jet and presents the results. The results are 
compared with the ones obtained with the previous works and the future expected bounds. 
Finally, section 4 concludes the paper.

\section{Effect of the anomalous couplings $\kappa$ and $\tilde{\kappa}$ on the cross section}

\begin{figure}
\centering
  \includegraphics[width=10cm,height=7cm]{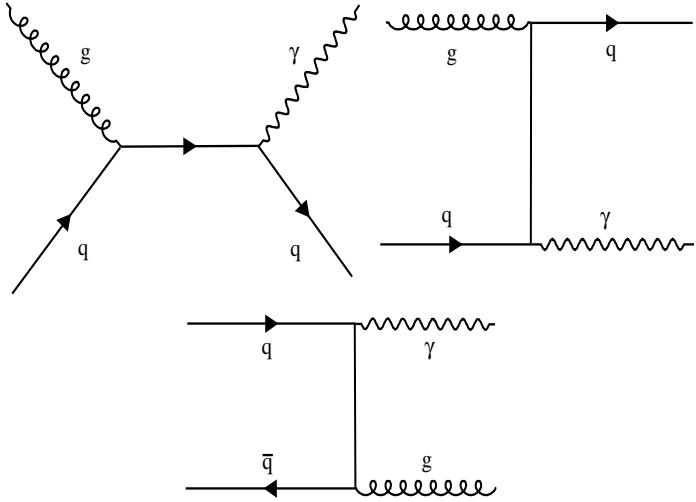}
   \caption{ The representative Feynman diagrams for production of $\gamma+jet$ at leading order 
in proton-proton collisions at the LHC.}
   \label{feynman}
\end{figure}

In this section we present the theoretical calculation of the $\gamma+jet$
production cross section at the LHC in the presence of chromoelectric and chromomagnetic
dipole moments of the quarks. 
The main contribution to $\gamma + jet$ final state in proton-proton collisions 
comes  when a hadronic jet and a photon is produced in a hard scattering. 
This can be achieved by the compton scattering of the quark gluon
$(gq\rightarrow \gamma q)$ and quark-antiquark annihilation $(q\bar{q} \rightarrow \gamma g)$ at leading order.
In Fig.\ref{feynman}, the representative Feynman diagrams for production of $\gamma+jet$ are depicted.
As mentioned previously,  the effective
interaction of $qqgg$ which is absent in the SM appears in the effective Lagrangian to ensure the gauge invariant \cite{wb,soni}.
This new four-point interaction does not affect the $\gamma+jet$ production.

Based on the effective Lagrangian introduced in Eq.\ref{eff}, the new Feynman rule
describing the interaction of a quark with gluon has the following form:
\begin{equation}
\label{qqgamma}
{\cal L}_{eff}={\cal L}_{SM}+{\cal L}_{q_i q_j g} =-g_{s} \bar q_j T^{a}_{ji} \left[ \gamma^\mu +\frac{i}{2m_q} \sigma^{\mu\nu}
 q_\nu (\kappa - i {\tilde{\kappa}} \gamma^5) 
\right ] q_i\; G^{a}_\mu \;,
\end{equation} 
where $q_i$ and $q_j$ are the quark spinors and $q_\nu$ is the four-momentum of the gluon.
It is notable that in general the anomalous couplings could be dependent on the gluon momentum transfer. 
However, the transfer momentum
is much smaller than the new physics scale therefore the dependency is neglected.
The color and spin averaged amplitude for $q(p) g(p') \to \gamma(k) q(k')$ is found as: 
\begin{equation}
\label{amp1}
\overline{\sum} |{M}|^2 = \frac{16\pi^2 \alpha_s
\alpha_{\rm em} e_q^2}{3} \biggr [ - \frac{{\hat{s}}^2 +{\hat{t}}^2}{\hat{s}\hat{t}} - \frac{\hat{u}}{2m_{q}^{2}}  
(\kappa^{2} +{\tilde{\kappa}}^{2} ) \biggr]
\end{equation}
where $s,t,u$ are the mandelstam variables defined as $\hat{s}=(p+p')^2\,, \;\; \hat{t}=(p-k)^2 \,, \;\; \hat{u}=(p-k')^2$
and $e_q$ is the electric charge of the quark $q$ in units of the proton charge.
The color and spin averaged amplitude for 
$q(p) \bar q (p') \to \gamma(k) g(k')$ has the following form:
\begin{equation}
\label{amp2}
\overline{\sum} |{M}|^2 = \frac{128\pi^2 \alpha_s
\alpha_{\rm em} e_q^2}{9} \biggr [ \frac{{\hat{t}}^2 +{\hat{u}}^2}{\hat{u}\hat{t}} + \frac{\hat{s}}{2m_{q}^{2}} 
(\kappa^{2} + {\tilde{\kappa}}^{2} ) \biggr] \;.
\end{equation}
All calculations are consistent with \cite{kd}. 
Now, the hadronic cross section is obtained by convoluting the parton level cross section with
the proton parton density functions:
\begin{eqnarray}
d\sigma(pp\rightarrow \gamma+jet) = \sum_{ij=qg}\int_{0}^{1} dx_{1} \int_{0}^{1} dx_{2} f_{i}(x_{1},Q^{2})f_{j}(x_{2},Q^{2})d\hat{\sigma}_{ij} 
\end{eqnarray}
where $f_i(x, Q^2)$ denotes the parton distribution functions (PDFs). We use an almost recently released CT10 \cite{ct10} set as for 
PDFs to calculate the cross section.
In this analysis, the nominal renormalization and factorization scales have
been set to the photon transverse momentum $\mu_{R}=\mu_{F}=Q=p_{T,\gamma}$.  
It should be mentioned here that there are higher order processes contributing to the total cross section
like $gg\rightarrow \gamma+g$.
The theoretical prediction for the higher order corrections leads to a k-factor of 1.3 \cite{nlo1,nlo2}. 
This is applied to include the next-to-leading order effects for $\gamma+jet$ cross section. 
We emphsize here that in reality  k-factor varies in different bins of the photon-jet invariant mass and is not fixed.
In \cite{nlo2}, the authors have calculated k-factor as a function of photon $p_{T}$ for Tevatron and for the LHC at the 
center-of-mass energy of 14 TeV. Since, the k-factor as a function of $M_{\gamma-jet}$ at the center-of-mass energy
of 8 TeV is not available we use the fixed value similar to other analyzes \cite{cms}. 

The dependence of the cross section to $\kappa$ and $\tilde{\kappa}$ is similar 
as the amplitudes of two subprocesses are proportional to $\kappa^{2} + {\tilde{\kappa}}^{2}$. 
Therefore, the effects because of nonzero CEDM is quite similar to nonzero CMDM in the total cross section.
If the $\gamma+jet$ total cross section will be measured at the LHC8 (LHC14) compatible with 
the SM prediction with a relative uncertainty of $5\%$,
an upper bound of $8.5\times 10^{-5}$ ($7 \times 10^{-5}$) on $\kappa$ or $\tilde{\kappa}$ will be obtained. 
However, in the next section we will show that the $\gamma+jet$ mass spectrum ($d\sigma/dM_{\gamma-jet}$) is more 
sensitive to CEDM and CMDM and gives stringent limits.

\section{Photon-jet mass spectrum sensitivity estimate }

In this section, we concentrate on the measured $\gamma+jet$ mass spectrum
to probe the chromoelectric and chromomagnetic dipole moments of the quarks.
In \cite{kd},  it has been shown that the prompt photon transverse momentum 
spectrum is sensitive to CEDM and CMDM in proton-antiproton collisions at the Tevatron.
Then using the CDF and D0 data, any value of $\kappa$ or $\tilde{\kappa}$ above 0.0027 has been excluded.

\begin{figure}
\centering
  \includegraphics[width=12cm,height=8cm]{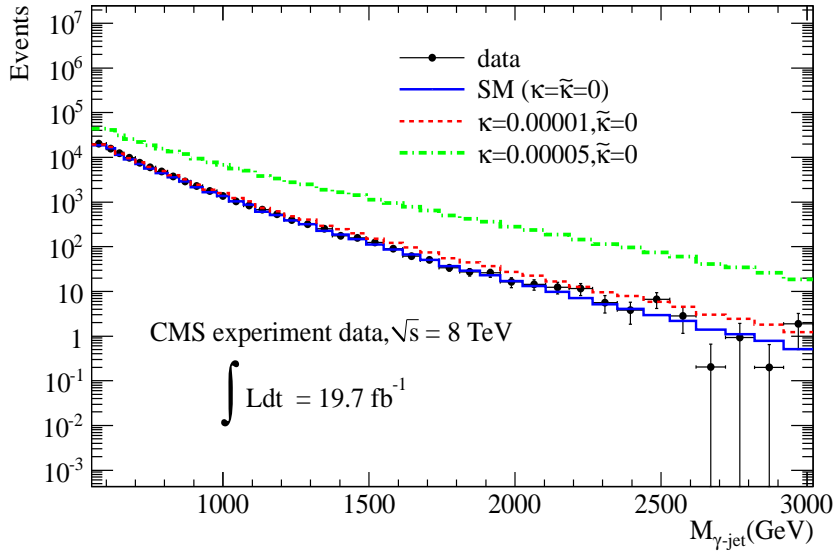}
   \caption{The measured and SM expected distributions of the invariant mass of $jet+\gamma$ at the LHC with the center-of-mass energy of 8 TeV. 
  The spectrum is weighted to the integrated luminosity of 19.7 fb$^{-1}$.  The expectation of the new effective Lagrangian for two values of 
 $\kappa$ is also presented.
}\label{mass}
\end{figure}

Recently, the CMS experiment has measured the $\gamma+jet$ 
mass spectrum in proton-proton collisions at the center-of-mass 
energy of 8 TeV with an integrated luminosity of 19.7 fb$^{-1}$ of data \cite{cms}.
Then the invariant mass spectrum has been used to look for signature of new physics (excited quarks)
after implementing the fiducial requirements on
both photon and jet. The mass spectrum is well fitted to a parameterization
which describes the SM prediction and no significant excess over the SM expectation has been observed. 
In Fig.\ref{mass}, the measured and SM expected distributions of the invariant mass of $\gamma+jet$ 
is shown after applying similar kinematic requirements as the CMS experiment.
The transverse momenta of photon and jet are required to be greater than 170 GeV. Photon
is restricted to be in the pseudorapidity range of $|\eta_{\gamma}|<1.44$ and jet is required 
to be in $|\eta_{jet}|<3$. The angular separation of jet and photon is required 
to be larger than 1.5 while the differences of pseudorapidities 
are required to be less than 2.0.
The effects of the presence of $\kappa$ or $\tilde{\kappa}$ is also shown in the plot.
As it can be seen, the photon-jet mass spectrum is affected by the presence of $\kappa$ or $\tilde{\kappa}$.
Nonzero values of  $\kappa$ or $\tilde{\kappa}$ leads to a significant increase in the cross section in particular in 
the high mass region. In this analysis, all quark flavors except for top quark are included in the calculations.
We have assumed that the CEDM and CMDM of these quarks are the same.  

\begin{figure}
\centering
  \includegraphics[width=8cm,height=6cm]{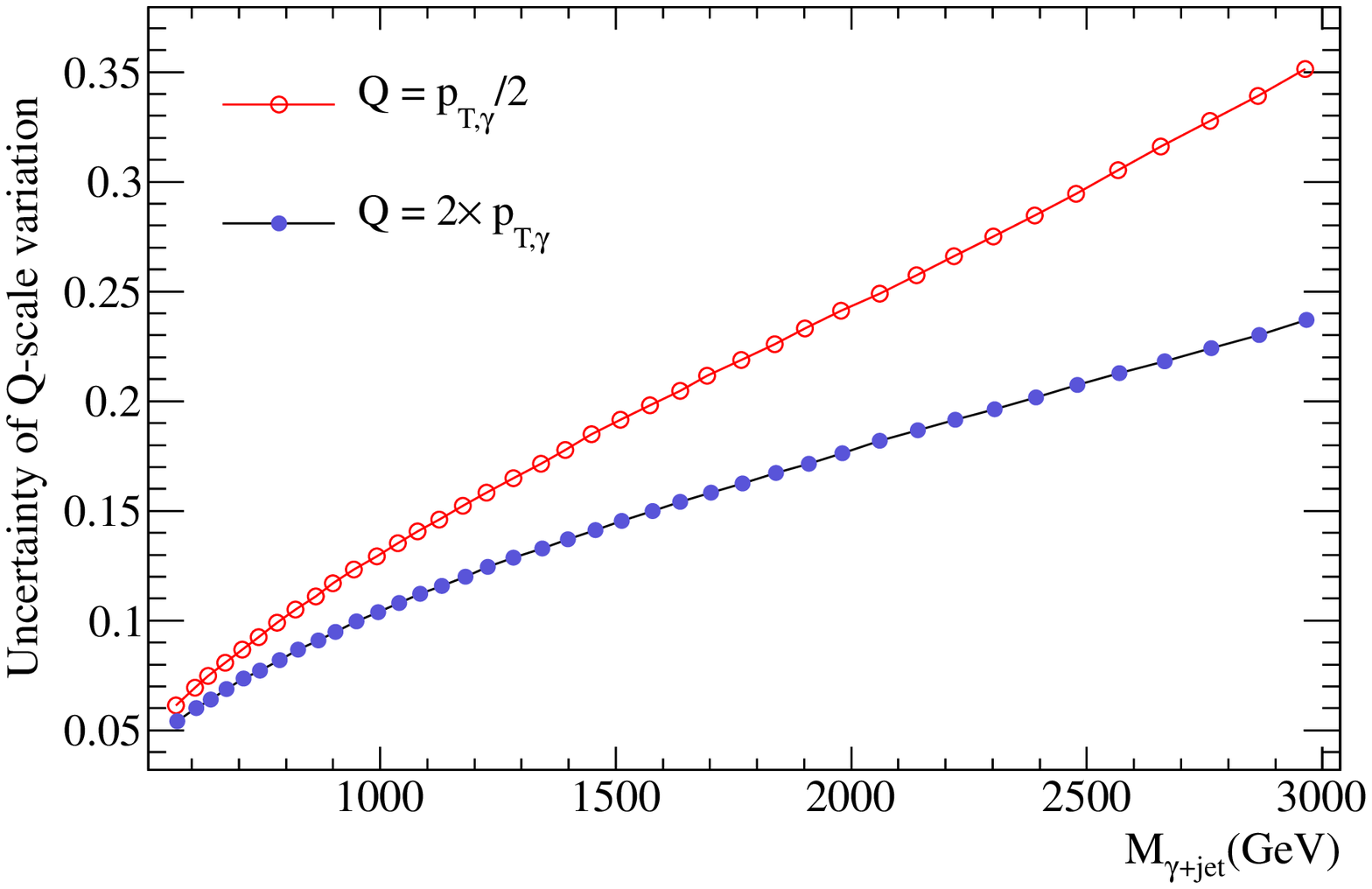}
  \includegraphics[width=8cm,height=6cm]{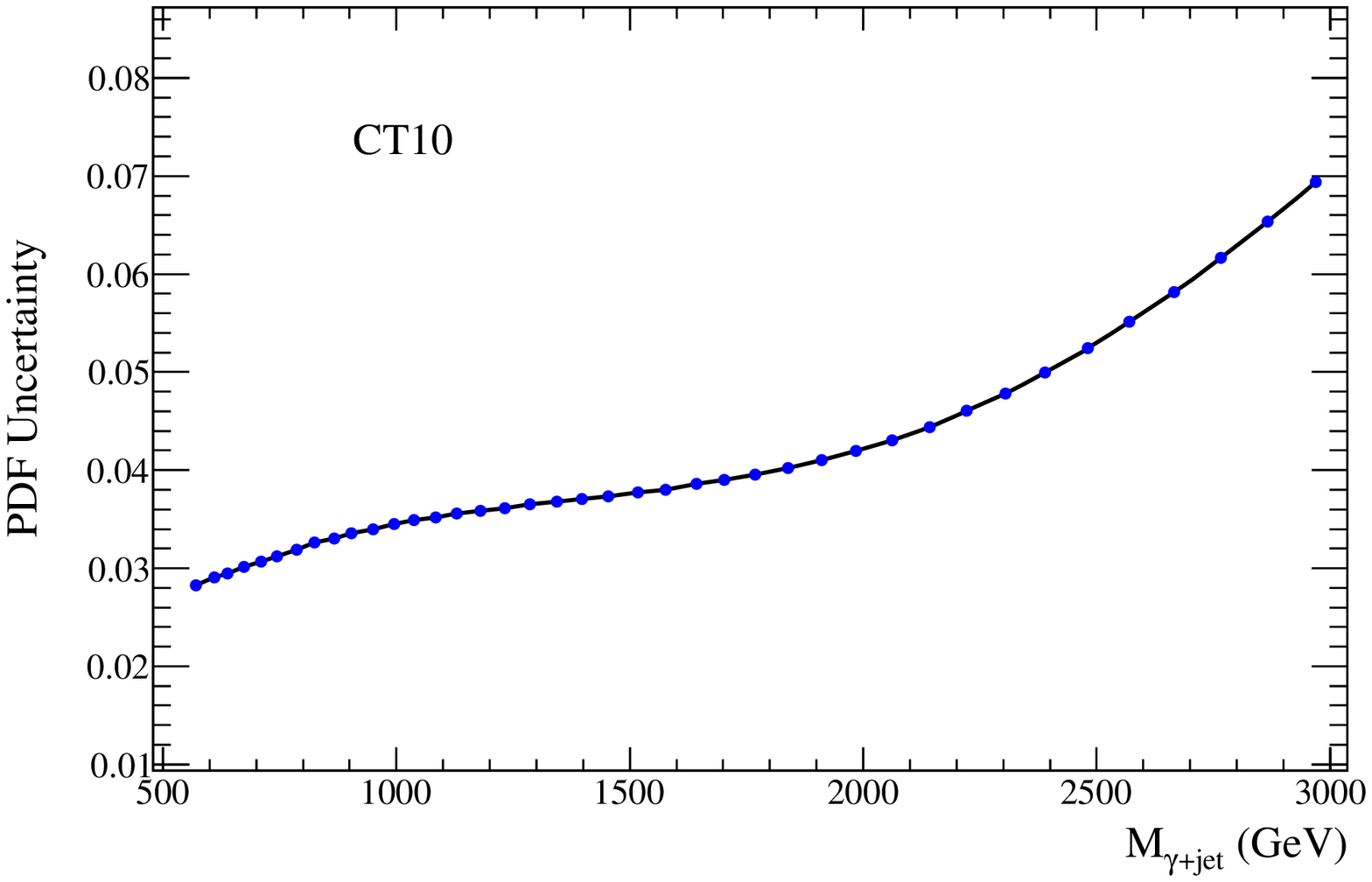}
   \caption{Left: the absolute value of the relative uncertainty due to variation of the factorization/renormalization scales.
   Right: the absolute value of the relative uncertainty due to choice of PDF obtained using the PDF4LHC recommendations.}
   \label{PDF}
\end{figure}

Now, in order to constrain the quark CEDM and CMDM, we combine the information of all bins of
photon-jet mass spectrum as shown in Fig.\ref{mass} into a global $\chi^{2}$
fit. The $\chi^{2}$ is defined as:
\begin{eqnarray}
\chi^{2} = \sum_{i=bins} \frac{(N_{i}^{obs}-N_{i}^{th})^2}{\Delta _{i}^2}
\end{eqnarray}
where the sum is over the bins of photon-jet invariant mass, $N_{i}^{th}$ is the number of expected events 
in a given theory, defined by the values of $\kappa$ and $\tilde{\kappa}$ in each mass bin, and 
$N_{i}^{obs}$ is the observed number of events in each bin. The uncertainty of the expectation in each bin
is denoted by $\Delta_{i}$ that covers both theoretical and experimental uncertainties. We have included 40
bins of the photon-jet invariant mass in $\chi^{2}$.
In order to have a realistic estimation of the upper limits on the anomalous couplings $\kappa$ and $\tilde{\kappa}$
all sources of uncertaities must be taken into account. In the denominator of $\chi^{2}$, $\Delta_{i}$ contains all the uncertainties.
There are several sources of uncertainties: statistical, systematic and theoretical uncertainties. The main sources of
theoretical uncertainties are the uncertainty due to variation of factorization/renormalization scales, the uncertainty 
originating from our limited knowledge of parton distribution functions, and the uncertainty on the value of strong coupling
constant $\alpha_{s}$.
To estimate the uncertainty from variation of factorization/renormalization scales, the scales are varied by a factor 
of 0.5 and 2. The aboslute value of the resulting differences with the nominal scale are shown in Fig.\ref{PDF} (left).
According to Fig.\ref{PDF}, the uncertainty increases with the photon-jet invariant mass. At the invariant mass
around 3 TeV, the uncertainty varies between $25\%-35\%$. In each bin, we take the average of uncertainty 
from $Q=2p_{T\gamma}$ and $Q=p_{T\gamma}/2$ variations.
 The uncertainty originating from the choice of parton distribution functions is calculated in bins of 
 photon-jet invariant mass according to the PDF4LHC recommendations \cite{pdflhc}. 
 The results are shown in the right side of Fig.\ref{PDF}.  The uncertainty varies from $3\%$
 to $7\%$ when the photon-jet invariant mass varies from 600 GeV to 3000 GeV. 
 For the sake of considering the uncertainty on the value of strong coupling constant $\alpha_{s}$,
 we vary the value of $\alpha_{s}$ around the nominal value $\alpha_{s}=0.118$ by $\pm 0.0002$. 
 The small dependence of strong coupling constant on $M_{\gamma-jet}$ is neglected.
There are several sources of instrumental uncertainties coming from jet energy and photon energy
resolutions. Conservatively, an overall value of $5\%$ in each bin of invariant mass is considered.
All uncertainties are considered as quadratic sum of each uncertainty in each bin of the 
photon-jet invariant mass: $\Delta^{2}=\sigma^{2}_{stat}+\sigma^{2}_{theory}+\sigma^{2}_{syst}$.

Now, we set upper limits on $\kappa$ or $\tilde{\kappa}$ at $95\%$ confidence level.
Similar to \cite{kd}, the results are also presented in terms of $\Lambda = \frac{2m_{q}}{\kappa}$.
Including only statistical uncertainties, the upper limit of $8.27 \times 10^{-6}$ is obtained. This upper limits gets looser
and reaches to $2.5\times 10^{-5}$ after considering all systematic and theoretical uncertainties.  This is an improvement
on the previous direct limits on CEDM or CMDM ($2.7 \times 10^{-3}$) with two order of magnitudes. 
We have put $\tilde{\kappa}=0$ and obtained upper limit on $\kappa$.  
Since the cross section is proportional to $\kappa^{2}+\tilde{\kappa}^{2}$, the same upper 
bound is obtained on $\tilde{\kappa}$.  In table \ref{tab}, the results are presented 
in terms of $\Lambda = \frac{2m_{q}}{\kappa}$ as well as $\kappa$.
The results are compared with the results that have been obtained in \cite{kd} for the future run of  LHC at 14 TeV 
center-of-mass energy. As it has been mentioned in the past, the analysis of \cite{kd} is based on the effect of 
CEDM and CMDM on the photon transverse momentum.

\section{Conclusions}

We have shown that the photon-jet invariant mass distribution
in proton-proton collisions at the LHC  receives significant contributions from the quarks chromoelectric ($\kappa$) and chromomagnetic 
($\tilde{\kappa}$) dipole moments in particular at large values of $\gamma+$jet invariant mass.
We use the measured $\gamma+$jet mass spectrum by the CMS experiment at the center-of-mass energy
of 8 TeV to derive upper limits on  quarks chromoelectric and chromomagnetic dipole moments.
All theoretical and systematic uncertainties are included in limit setting.  We have calculated the 
uncertainty originating from variation of factorization/renormalization scales as a function of
the photon-jet invariant mass. The uncertainty from variation of factorization/renormalization scales
increases with increasing the photon-jet invariant mass and reaches up to $35\%$ at invariant mass 
of 3 TeV. The uncertainty due to choice of PDF has been calculated using the PDF4LHC recommendations.
We exclude $\kappa$ or $\tilde{\kappa}$ above
$10^{-5}$ at 95$\%$ confidence level. The sensitivity is also presented in terms of an energy scale
parameter $\Lambda=2m_{q}/\kappa$. 
Any value of $\Lambda$ below 24 TeV has been excluded using 19.7 fb$^{-1}$ of LHC data at the 
center-of-mass energy of 8 TeV.

\begin{table}\caption{Comparison of upper limits on $\kappa$ or $\tilde{\kappa}$ and also on $\Lambda$ from the this analysis and from \cite{kd}.}
\begin{center}
\begin{tabular}{c|cccc}
\hline \hline
     Experiment      &  $\sqrt{s}$ (TeV)    &     $\int \mathcal{L}$ (fb$^{-1}$)  &  limit on $\kappa$                 &              $\Lambda$  (TeV)   \\  \hline \hline  
     Tevatron($p_{T\gamma}$)         &  1.8               &          0.1                                      &        $2.7\times 10^{-3}$      &               0.7            \\ 
           LHC($p_{T\gamma}$)          &    14              &           10                                      &        $1.3\times 10^{-4}$       &               4.5             \\
           LHC($p_{T\gamma}$)          &    14              &           100                                      &     $9.5\times 10^{-5}$         &              6.3           \\  \hline
           LHC($M_{\gamma-jet}$)          &     8               &            19.7                                   &       $2.5\times 10^{-5}$        &              24                    \\  \hline \hline
\end{tabular}\label{tab}
\end{center}
\end{table}

\end{document}